\documentstyle[12pt]{article}

\input epsf

\begin{document}
\baselineskip=20.5pt

\def\beqra{\begin{eqnarray}} \def\eeqra{\end{eqnarray}}
\def\beqast{\begin{eqnarray*}}
\def\eeqast{\end{eqnarray*}}
\def\beq{\begin{equation}}      \def\eeq{\end{equation}}
\def\be{\begin{enumerate}}   \def\ee{\end{enumerate}}

\def\fnote#1#2{\begingroup\def\thefootnote{#1}\footnote{
#2}
\addtocounter
{footnote}{-1}\endgroup}

\def\itp#1#2{\hfill{NSF-ITP-{#1}-{#2}}}

\def\gam{\gamma}
\def\Gam{\Gamma}
\def\la{\lambda}
\def\eps{\epsilon}
\def\La{\Lambda}
\def\si{\sigma}
\def\Si{\Sigma}
\def\al{\alpha}
\def\Tha{\Theta}
\def\tha{\theta}
\def\vphi{\varphi}
\def\del{\delta}
\def\Del{\Delta}
\def\ab{\alpha\beta}
\def\om{\omega}
\def\Om{\Omega}
\def\mn{\mu\nu}
\def\mun{^{\mu}{}_{\nu}}
\def\kap{\kappa}
\def\rsi{\rho\sigma}
\def\beal{\beta\alpha}

\def\til{\tilde}
\def\rta{\rightarrow}
\def\eqv{\equiv}
\def\nab{\nabla}
\def\pa{\partial}
\def\sit{\tilde\sigma}
\def\ul{\underline}
\def\indt{\parindent2.5em}
\def\nd{\noindent}

\def\rsi{\rho\sigma}
\def\beal{\beta\alpha}

\def\caa{{\cal A}}
\def\cb{{\cal B}}
\def\cac{{\cal C}}
\def\cd{{\cal D}}
\def\ce{{\cal E}}
\def\cf{{\cal F}}
\def\cg{{\cal G}}
\def\cah{{\cal H}}
\def\ci{{\cal I}}
\def\cj{{\cal{J}}}
\def\ck{{\cal K}}
\def\cl{{\cal L}}
\def\cm{{\cal M}}
\def\cn{{\cal N}}
\def\cO{{\cal O}}
\def\cp{{\cal P}}
\def\car{{\cal R}}
\def\cs{{\cal S}}
\def\ct{{\cal{T}}}
\def\cu{{\cal{U}}}
\def\cv{{\cal{V}}}
\def\cw{{\cal{W}}}
\def\cx{{\cal{X}}}
\def\cy{{\cal{Y}}}
\def\cz{{\cal{Z}}}

\def\raisenot{\raise .5mm\hbox{/}}
\def\nota{\ \hbox{{$a$}\kern-.49em\hbox{/}}}
\def\notA{\hbox{{$A$}\kern-.54em\hbox{\raisenot}}}
\def\notb{\ \hbox{{$b$}\kern-.47em\hbox{/}}}
\def\notB{\ \hbox{{$B$}\kern-.60em\hbox{\raisenot}}}
\def\notc{\ \hbox{{$c$}\kern-.45em\hbox{/}}}
\def\notd{\ \hbox{{$d$}\kern-.53em\hbox{/}}}
\def\notbd{\ \hbox{{$D$}\kern-.61em\hbox{\raisenot}}} 
\def\note{\ \hbox{{$e$}\kern-.47em\hbox{/}}}
\def\notk{\ \hbox{{$k$}\kern-.51em\hbox{/}}}
\def\notp{\ \hbox{{$p$}\kern-.43em\hbox{/}}}
\def\notq{\ \hbox{{$q$}\kern-.47em\hbox{/}}}
\def\notW{\ \hbox{{$W$}\kern-.75em\hbox{\raisenot}}}
\def\notz{\ \hbox{{$Z$}\kern-.61em\hbox{\raisenot}}}
\def\notpa{\hbox{{$\partial$}\kern-.54em\hbox{\raisenot}}}

\def\fo{\hbox{{1}\kern-.25em\hbox{l}}}  
\def\rf#1{$^{#1}$}
\def\bx{\Box}
\def\tr{{\rm Tr}}
\def\rmtr{{\rm tr}}
\def\dgg{\dagger}

\def\lag{\langle}
\def\rag{\rangle}
\def\bmid{\big|}

\def\vlap{\overrightarrow{\La p}} 
\def\lrta{\longrightarrow}
\def\lrar{\raisebox{.8ex}{$\longrightarrow$}}
\def\rlarw{\longleftarrow\!\!\!\!\!\!\!\!\!\!\!\lrar}

\def\llra{\relbar\joinrel\longrightarrow}     
\def\mapright#1{\smash{\mathop{\llra}\limits_{#1}}}
\def\mapup#1{\smash{\mathop{\llra}\limits^{#1}}}
\def\asymptotic{{_{\stackrel{\displaystyle\longrightarrow}
{x\rightarrow\pm\infty}}\,\, }} 
\def\asymptext{\raisebox{.6ex}{${_{\stackrel{\displaystyle\longrightarrow}
{x\rightarrow\pm\infty}}\,\, }$}} 

\def\7#1#2{\mathop{\null#2}\limits^{#1}}   
\def\5#1#2{\mathop{\null#2}\limits_{#1}}   
\def\too#1{\stackrel{#1}{\to}}
\def\tooo#1{\stackrel{#1}{\longleftarrow}}
\def\nout{{\rm in \atop out}}

\def\one{\raisebox{.5ex}{1}}
\def\BM#1{\mbox{\boldmath{$#1$}}}

\def\ltsim{\matrix{<\cr\noalign{\vskip-7pt}\sim\cr}}
\def\gtsim{\matrix{>\cr\noalign{\vskip-7pt}\sim\cr}}
\def\haf{\frac{1}{2}}


\def\place#1#2#3{\vbox to0pt{\kern-\parskip\kern-7pt
                             \kern-#2truein\hbox{\kern#1truein #3}
                             \vss}\nointerlineskip}

\def\illustration #1 by #2 (#3){\vbox to #2{\hrule width #1
height 0pt
depth
0pt
                                       \vfill\special{illustration #3}}}

\def\scaledillustration #1 by #2 (#3 scaled #4){{\dimen0=#1
\dimen1=#2
           \divide\dimen0 by 1000 \multiply\dimen0 by #4
            \divide\dimen1 by 1000 \multiply\dimen1 by #4
            \illustration \dimen0 by \dimen1 (#3 scaled #4)}}

\def\ON{{\cal O}(N)}
\def\UN{{\cal U}(N)}
\def\bdPh{\mbox{\boldmath{$\dot{\!\Phi}$}}}
\def\bPh{\mbox{\boldmath{$\Phi$}}}
\def\bPhs{\bPh^2}
\def\sef{S_{eff}[\sigma,\pi]}
\def\sigx{\sigma(x)}
\def\pix{\pi(x)}
\def\bph{\mbox{\boldmath{$\phi$}}}
\def\bphs{\bph^2}
\def\ex{\BM{x}}
\def\exs{\ex^2}
\def\xdot{\dot{\!\ex}}
\def\y{\BM{y}}
\def\ys{\y^2}
\def\ydot{\dot{\!\y}}
\def\pat{\pa_t}
\def\pax{\pa_x}

\renewcommand{\theequation}{\arabic{equation}}


\itp{97}{073}\\

\hfill{cond-mat/9706218}\\
\vspace*{.3in}
\begin{center}
 \large{\bf Non-Hermitean Localization and De-Localization}
\normalsize

\vspace{36pt}
Joshua Feinberg\fnote{}{{\it e-mail: joshua, zee@itp.ucsb.edu}}
 \& A. Zee$^{}$\\

\vspace{12pt}
 {\small \em Institute for Theoretical Physics,}\\ {\small \em
University of California, Santa Barbara, CA 93106, USA}
\vspace{.6cm}

\end{center}

\begin{minipage}{5.3in}
{\abstract~~~~~We study localization and delocalization in a class of non-hermitean Hamiltonians inspired by the problem of vortex pinning in superconductors. In various simplified models we are able to obtain analytic descriptions, in particular of the non-perturbative emergence of a forked structure (the appearance of ``wings") in the density of states. We calculate how the localization length diverges at the localization-delocalization transition. We map some versions of this problem onto a random walker 
problem in two dimensions. For a certain model, we find an intricate structure in its density of states.}
\end{minipage}

\vspace{48pt}


\vfill
\pagebreak

\setcounter{page}{1}

\section{Introduction}

Non-hermitean random matrix theory \cite{recent} has been applied recently to a number of interesting physical situations. An interesting issue among these physical situations is the study of localization-delocalization transitions in non-hermitean random Hamiltonians. The earliest discussion of localization in the context of a non-hermitean hydrodynamics problem appears to be in \cite{miller}. Later on, Hatano and Nelson \cite{hatano} have mapped the 
problem of the vortex line pinning in superconductors to a problem involving a non-hermitean random Hamiltonian. When a current is passed through a superconductor, vortex lines tend to drift in a direction perpendicular to the current, but this tendency is counteracted by
impurities on which the vortex lines are pinned. It is expected that at
some critical current the vortex lines become unpinned or delocalized. In
the simplest model of this problem, the physics is modelled by a quantum
particle hopping on a ring, whose rightward (or counterclockwise) hopping
amplitude $t e^h /2$ is different from its leftward hopping amplitude $t
e^{-h} /2$. Note that $ih$ may be thought of as an imaginary gauge field.
On each site of the ring is a random potential $w_i$ which
tries to trap the particle. The Hamiltonian 
\beq\label{h}
H=H_0+W
\eeq
is thus the sum of 
the deterministic non-hermitean hopping term 
\beq\label{h0}
H_{0 ij} ={t\over 2}\left( e^h~\delta_{i+1, j} + e^{-h}~\delta_{i, j+1}\right)\quad , \quad i,j = 1, \cdots N
\eeq
(with the obvious periodic identification $i+N\equiv i$ of site indices) and 
the hermitean random potential term  
\beq\label{W}
W_{ij} = w_i~\delta_{i, j}\,.
\eeq
The number of sites $N$ is understood to be tending to infinity. This
problem has been studied by a number of authors \cite{efetov,feinzee,zahed, beenakker} following Hatano and Nelson. Note that the Hamiltonian not only breaks 
parity as expected but is non-hermitean, and thus has complex eigenvalues. It 
is represented by a real non-symmetric matrix, with the reality implying 
that if $E$ is an eigenvalue, then $E^*$ is also an eigenvalue.

With no impurities ($w_i=0$) the Hamiltonian is immediately solvable by
Bloch's theorem with the eigenvalues
\beq\label{spectrum}
E_n = t~{\rm cos}~( {2\pi n\over N} - ih)\,, \quad (n = 0, 1, \cdots, N-1)\,, 
\eeq
tracing out an ellipse. The corresponding wave functions $\psi^{(n)}_j\sim {\rm exp}~2\pi i n j/N$ are obviously extended. Note that in the limit of zero non-hermiticity
($h=0$) the ellipse collapses to a segment on the real axis as expected.
With impurities present the spectrum can be obtained by numerical work as
was done extensively by Hatano and Nelson and as is shown in Fig. 1. 

\begin{figure}
\epsfysize=4 truein
\epsfxsize=4 truein
\centerline{\epsffile{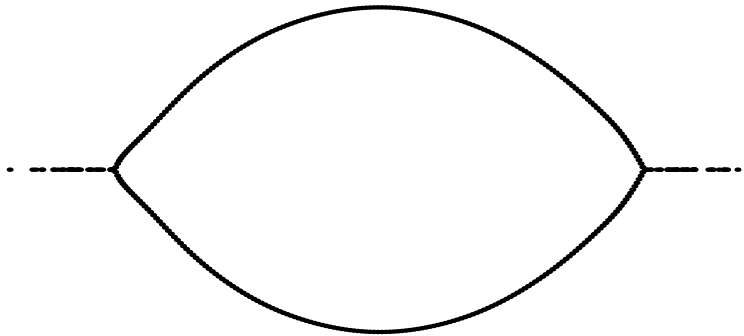}}
\vspace{-0.1 truein}
\caption[]{The spectrum of the Hamiltonian in Eq. (\ref{h}) for $h=0.5$, 
$t=2$ and $N=400$ sites. Shown here is the spectrum for one particular realization of site energies taken from a flat distribution with $-2\leq w_i\leq 2$. A finite fraction of eigenvalues has clearly snapped onto the real axis. (The coordinate axes have been suppressed to make the appearance of the snapped eigenvalues clearer.)}
\end{figure}
Two ``wings" have emerged out of the two ends of the ellipse. Some eigenvalues
have become real. Evidently, the ``forks" where the two wings emerge out of
the ellipse represent a non-perturbative effect, and cannot be obtained by
treating the impurities perturbatively. It is thus something of a challenge
to obtain the two wings analytically. In this paper, we address this and
other problems.

As discussed in Section 6 of \cite{feinzee} this behavior could be understood qualitatively by a simple example. In ordinary hermitean quantum mechanics, it is a familiar textbook dictum that nearby eigenvalues repel. In contrast, two nearby eigenvalues in the complex plane, separated along the imaginary direction, attract each other under a hermitean perturbation. To see this consider the $2\times 2$ matrix 
\beqast
\left(\begin{array}{cc} i\epsilon & 0\\{} & {}\\ 0 & -i\epsilon\end{array}\right) +\left(\begin{array}{cc} 0 & w\\{} & {}\\ w & 0\end{array}\right)
\eeqast
with eigenvalues are $\pm i \sqrt{\epsilon^2-w^2}$. Thus, for $w<\epsilon$, the original eigenvalues $\pm i\epsilon$ attract each other, but remain  on the imaginary axis. However, as soon as $w\geq\epsilon$, they snap onto the real axis and start repelling each other. Let us start with the ellipse in the 
absence of impurities. Near each tip of the ellipse, there is a pair of eigenvalues separated slightly along the imaginary direction and lying on opposite sides of the real axis. They attract each other and thus approach the real axis, but as soon as the two ``friends" arrive on the real axis, they immediately repel each other
(as well as the eigenvalue already on the real axis.) Obviously, this
process repeats itself with the next pair of eigenvalues, and thus leads to
the formation of the wings.

Hatano and Nelson emphasized that $H_0$ has a special property, namely that 
by a (non-unitary) gauge transformation, all the non-hermiticity can be concentrated on one arbitrarily chosen bond, without changing the spectrum. This leads to an extraordinarily simple argument that the states
corresponding to complex eigenvalues are extended, that is, delocalized.
Let $H\psi=E\psi$. Assume that $\psi$ is localized around some site $j$. We can always gauge the non-hermiticity to a link which is located arbitrarily
far away from the site $j$, where $|\psi|$ is exponentially small. Thus, if 
we cut the ring open at that link, the effect on the Schr\"odinger equation would be exponentially small, and would vanish completely in the limit $N\rightarrow\infty$. It follows from 
this simple argument that if we replace the periodic boundary condition in solving $H\psi=E\psi$ by an open chain boundary condition, the localized 
part of the spectrum of (\ref{h}) would not be affected. But for $H$ with an open chain boundary condition, the gauge transformation just mentioned
may be used to gauge away the non-hermiticity completely, meaning that the Hamiltonian is in effect hermitean with real eigenvalues. 
We thus conclude that all localized eigenstates of (\ref{h}) correspond to real eigenvalues (in the large $N$ limit.) In other words, the states
corresponding to complex eigenvalues are extended, that is, delocalized.

Remarkably, non-hermitean localization theory is simpler in this respect
than the standard hermitean localization theory of Anderson and others \cite{anderson}. To understand the localization transition, one has to study only the density of eigenvalues, or equivalently, the one-point Green's function, rather than the two-point Green's function.

\pagebreak

\section{The single impurity case}

In this section our philosophy is to find the simplest version of the hopping model described in the Introduction which we can solve exactly, but yet manages to capture the essential physics involved, including the
non-perturbative emergence of the two ``wings" along the real axis. The desired simplification is to replace the $N$ random impurities in (\ref{W}) by a single impurity. With this simplification we do not need to specify the precise form of the non-hermitean $H_0$ beyond assuming that it is translationally invariant.
We thus replace the $W_{i,j}$ in (\ref{W}) by
\beq\label{w1}
W_{i,j} = w\,\delta_{i,1}\,\delta_{j,1}
\eeq
where $w$ is drawn from some probability distribution $P(w)$. 

Our task is to calculate the averaged Green's function  
\beq\label{greens}
G(z,z^*) = \langle {1\over N}~{\rm tr}~{1\over z-H_0-W}\rangle
\eeq
of the random Hamiltonian $H=H_0+W$, from which we may calculate
the eigenvalue density of $H$. 

Away from the location of the spectrum of $H_0$ in the complex $z$ plane
we expand (\ref{greens}) in powers 
of $1/(z-H_0)$, and thus obtain 
\beq\label{intermediate}
G(z) = G_0(z) - {1\over N} \left[ {\pa\over \pa z}\left({1\over z-H_0}\right)_{1,1}\right]\sum_{k=1}^{\infty}\langle w^k\rangle \left[\left({1\over z-H_0}\right)_{1,1}\right]^{k-1}
\eeq
where 
\beq\label{g0}
G_0(z) = \langle {1\over N}~{\rm tr}~{1\over z-H_0}\rangle
\eeq
is the Green's function of $H_0$. Due to the translational invariance of $H_0$
we have  
\beqast
\left({1\over z-H_0}\right)_{1,1} = G_0(z)
\eeqast
and so 
\beqra\label{gz1}
G(z) &=& G_0(z) - {1\over N}~{\pa G_0(z)\over \pa z}\sum_{k=1}^{\infty}\langle w^k\rangle \left[G_0(z)\right]^{k-1}\nonumber\\
&=& G_0(z) - {1\over N}~{\pa G_0(z)\over\pa z}~\langle {w\over 1-wG_0(z)} \rangle \,.
\eeqra
Observe that the effect of the single impurity on the Green's function is of order $1/N$, as should be expected.

We stress again that up to this point we did not adhere to any specific translationally invariant $H_0$ nor did we specify any particular probability distribution $P(w)$ in our derivation of (\ref{gz1}). We also note that our derivation is exact for any value of $N$.

For finite $N$ the singularities of $G_0(z)$ are isolated simple poles located at the eigenvalues of $H_0$. The effect of the single impurity on any of these poles would be to move it around in an amount which depends on the typical
scale $r$ of the distribution $P(w)$. Thus, if $z=z_0$ is one of the poles of $G_0(z)$, we expect that for small values of the scale $r$, the full Green's
function $G(z)$ will also have a pole near $z=z_0$. Thus, in the vicinity of $z=z_0$ (ignoring all the other poles) we may approximate
\beq\label{g0pole}
G_0(z)\sim {1\over N} {1\over z-z_0}\,.
\eeq
Substitution of (\ref{g0pole}) into (\ref{gz1}) yields the simple result 
\beq\label{gz3}
G(z)\sim {1\over N} \langle {1\over z-z_0-{w\over N}}\rangle\,.
\eeq
Eq. (\ref{gz3}) is nothing but the result of first order perturbation
theory (after all, we ignored the effect of all states other the eigenstate 
associated with $z_0$.)

We can now understand the non-perturbative emergence of wings under very
general circumstances. Evidently, besides the ``trivial" poles just
mentioned, $G(z)$ also has a pole whenever $wG_0(z)=1$ with $w$ in the support of $P(w)$. We have  
\beq\label{g0gen}
G_0(z,z^*) = \int~d^2x'~ {\rho_0(x',y')\over z-(x'+iy')}
\eeq
where $\rho_0(x,y)\equiv (1/N) \sum_i~\delta(x-{\rm Re}~E_i)~\delta(y-{\rm Im}~
E_i)$ is the density of eigenvalues $E_i$ of $H_0$.  All we require for the
following discussion is that $\rho_0(x,y) = \rho_0(x,-y)$.  This is true for 
all the Hamiltonians $H_0$ considered in this paper.  Then on the (positive) real axis $G_0(x)$ is real, and decreasing
outside the spectrum of $H_0$.  (Indeed, it is well known that we can interpret
the real and imaginary parts of $G_0(z,z^*)$ as the electrostatic field
$\vec E = (E_x, E_y) = ({\rm Re}~G_0, -{\rm Im}~G_0)$ generated by the charge density $\rho_0$.)  Thus, if the real quantity $1/w$ lies between  $G_0(x_{{\rm edge}})$ and zero (where $x_{{\rm edge}}$ denotes the intersection of the edge 
of $\rho_0$  with the real
axis), we will have a pole on the real axis at some $x_{*}(w)$.  Averaging over $w$ we thus obtain a wing on the positive real axis.  A similar discussion can
obviously be given for the negative real axis.  It is also clear that there
is no solution of $wG_0(z)=1$  for $z$ outside $\rho_0$  and away from the real axis.

The probability distribution $P(w)$ will in general depend on some set of
parameters $\{r_i\}$ and for some given $\{r_i\}$ it is of course possible that $1/w$ does not
lie between $G_0(x_{{\rm edge}})$ and zero.  Thus, for some critical values   $\{r_i^c\}$ there will be a
transition at which the wings, and hence the localized states associated
with them, disappear.

In the $N\rightarrow\infty$ limit the eigenvalues of $H_0$ become 
dense and will either trace out a curve in the complex energy plane 
(analogous to the ellipse associated with (\ref{spectrum})), or fill out a two dimensional region (as for example, in two dimensional hopping problems.)
We now focus our
attention at the hopping Hamiltonian $H_0$ in (\ref{h0}).  To make things as simple as possible we let the parameters in (\ref{h0}) tend to the (maximally non-hermitean)  
limit $h\rightarrow\infty$ and $t\rightarrow 0$ such that 
\beq\label{thlim}
t~e^h\rightarrow 2
\eeq
(and obviously $t~e^{-h}\rightarrow 0$). In this limit (\ref{spectrum}) changes into 
\beq\label{circle}
E_n = {\rm exp}~{2\pi i n\over N}\,, \quad (n = 0, 1, \cdots, N-1)\,,
\eeq
and the ellipse associated with (\ref{spectrum}) expands into the unit 
circle. Furthermore, the full Schr\"odinger equation $(H_0+W)\psi=E\psi$ 
becomes simply
\beq\label{schrodinger}
w\psi_1+\psi_2 = E\psi_1\,,\quad \psi_i=E\psi_{i-1}\,.
\eeq
As a consequence, 
\beq\label{averatio}
\langle {\psi_i\over\psi_{i-1}} \rangle = \langle E\rangle
\eeq
from which we extract the localization length of the state $\psi$ 
simply as
\beq\label{loc}
L(E)\sim  {1\over {\rm log} \Big | \langle E\rangle\Big |} \,.
\eeq
At the points where the wings join onto the circle the localization length diverges as expected. 

The Green's function associated with (\ref{circle})
\beqast
G_0(z) = {1\over N}~\sum_{n=0}^{N-1}
~{1\over z-{\rm exp}~{2\pi i n \over N}}
\eeqast
may be approximated in the large $N$ limit by 
\beqast
G_0(z) = \oint {dw\over 2\pi i} {1\over w(z-w)} = \left\{\begin{array}{c}~0\,,\quad~ |z|<1\\~1/z\,,\quad |z|>1\,.\end{array}\right.
\eeqast
Substituting this expression into (\ref{gz1}) we obtain 
\beq\label{gcirc}
G(z) = \left\{\begin{array}{c}~0\,,\quad\quad\quad\quad\quad\quad |z|<1\\{}\\~{1\over z}+{1\over 2Nz}\langle{w\over z-w}\rangle\,,\quad |z|>1\,.\end{array}\right.
\eeq
Note that independently of $P(w)$ the disk inside the unit circle remains devoid of eigenvalues. To get a hold of the correction $G(z)-(1/z)$ in the outer region it is instructive to carry out some explicit calculations with
particular probability distributions $P(w)$. 

As our first concrete example, we take 
\beq\label{pw1}
P(w) = {1\over 2} [\delta (w-r) + \delta (w+r)]
\eeq
with some scale $r$. We then find from (\ref{gcirc}) that 
\beq\label{gz2}
G(z) = {1\over z} + {r^2\over 2Nz^2} \left({1\over z-r} + {1\over z+r}\right)\,,\quad |z|>1\,.
\eeq
If $r>1$, then $G(z)$ has two new poles on 
the real axis at $z=\pm r$, each with a residue $1/2N$. The critical value for $r$ to induce these two new poles is $r_c=1$. The existence of this 
critical value is a non-perturbative phenomenon (though its particular 
numerical value is retrospectively not surprising at all, being set by (\ref{circle}).)  From (\ref{loc}) we find that the localization length of the states associated with these poles is $L(r)\sim 1/{\rm log}~r$, and they thus become extended as $r\rightarrow r_c=1$.

Consider next the box distribution 
\beq\label{box}
P(w) = {1\over 2V}~\theta (V^2-w^2)
\eeq
for which 
\beq\label{circbox}
G(z) = {1\over z} - {1\over 2NV}~{\rm log} {z-V\over z+V}\,,\quad |z|>1\,.
\eeq
The eigenvalue density $\rho(x,y)$ (with $z=x+iy$) is related to the Green's function by the general relation \cite{feinzee}
\beq\label{generalrelation}
\rho(x,y) = {1\over\pi}~{\pa\over \pa z^*} G(z,z^*)
\eeq
which gives\footnote{Here we also use
the fact that in the cut plane (with the cut running along the negative real axis) $(\pa/\pa z^*) {\rm log} z \equiv (1/2)(\pa_x + i\pa_y) {\rm  log} z = -\pi\theta(-x)\delta (y)$}
\beq\label{rhobox}
\rho(x,y) = {1\over 2NV} \delta(y) \theta(V-|x|)\,,\quad |z|=|x|>1\,.
\eeq
As in the previous example, non-perturbative effects generate a 
critical value for $V$, namely $V_c=1$. For $V>V_c$, the density
of eigenvalues develops two symmetric wings of length $V-V_c$ such that 
the fraction of eigenvalues residing in these wings is ${1\over N}(1-{1\over V})\,.$ The localization length $L(E)$ of states that reside in the wings is finite for $|E|>V_c$ and diverges logarithmically
as $|E|\rightarrow V_c$.

As our final example for this section we consider the Lorentzian distribution
\beq\label{lorentz}
P(w) = {\gam\over \pi} {1\over w^2+\gam^2}\,,
\eeq
with its long tails extending to infinity.
The fraction of realizations in which an impurity potential is stronger than 
the unit scale set by (\ref{circle}) is $(1/\pi)~{\rm arctan}~\gam$.
We find that outside the unit circle
\beq\label{glorentz}
G(z) = {1\over z} -{i\gam\over Nz}\left[{1\over z+i\gam}~\theta({\rm Im} ~z)-{1\over z-i\gam}~\theta(-{\rm Im}~ z)\right]\,,\quad |z|>1
\eeq
and thus using (\ref{generalrelation}) we find that as in the previous example, the density of eigenvalues develops wings along the real energy axis
given by 
\beq\label{rholorentz}
\rho_{\rm wings}(x,y) = {1\over N\pi}~{\gam \over x^2+\gam^2} ~ \theta (x^2-1)~\delta(y)\,.
\eeq
Due to the long tails of (\ref{lorentz}), the wings extend to infinity, and the fraction of states that reside in them is $(1/N\pi)\,{\rm arctan}~\gam$, namely, the fraction of ``strong" impurity realizations
divided by $N$. Note also that there is no critical value for $\gam$, namely, the wings appear for all values of $\gam$.

Note that to obtain the analog of (\ref{gz1}) in the case of two impurities already involves a non-trivial combinatorial calculation involving $G_{ii}, G_{jj}$, and $G_{ij}$, where $i$ and $j$ denote the locations of the two impurities. (It is clear though, that the results of this section as they stand, are still applicable to the problem with many impurities, as long as the separation between impurities is larger than the localization length.) Remarkably, however,
we show in the next two sections that taking the maximally non-hermitean limit (\ref{thlim}) which restrict the particle to ``one way" hopping, we can readily treat the generic problem of many impurities.

Finally, we stress that our derivation of (\ref{gz1}) was not limited to one
dimensional hopping. Given $G_0(z)$ for a translationally invariant $H_0$ 
in any number of dimensions, $G(z)$ as given by (\ref{gz1}) is still valid outside the support of the spectrum of $H_0$. In general, the spectrum
of $H_0$ in higher dimensions will fill a two dimensional region in the complex
energy plane. In order to calculate $G(z,z^*)$ in that region, and in fact, to see how that region is affected by the impurity, we might have to resort 
to the method of Hermitization discussed in \cite{feinzee}.
\pagebreak

\section{Maximal non-hermiticity and many impurities: ``One Way" models}
In the preceding section, our exact result (\ref{gz1}) was obtained for any value of the parameter $h$. We may perhaps hope that we can study the many impurities problem in the particularly symmetric case provided by the limit (\ref{thlim}),
namely $t~e^h\rightarrow 2\,,\,t~e^{-h}\rightarrow 0$, in which the ellipse of eigenvalues becomes the circle (\ref{circle}). Thus, we will study the case of
maximal non-hermiticity in which $H$ is given by
\beq\label{maxnh}
H_{ij} = \delta_{i+1, j} + w_i~\delta_{i, j}\quad , \quad i,j = 1, \cdots N
\eeq
(with the obvious periodic identification $i+N\equiv i$ of site indices.) We refer to this class of models in which the particle only hops one way, as ``one way models."

We have found a particularly simple example in which 
\beq\label{pw}
P(w_i) = {1\over 2} [\delta (w_i-r) + \delta (w_i+r)]\,. 
\eeq
We now proceed to calculate the density of eigenvalues $E_i$ in this ``one way sign model", using the master formula
\beqra\label{master}
\rho(x,y)&\equiv&\langle {1\over N} \sum_i \delta(x-{\rm Re} E_i)~ \delta (y- {\rm Im}
E_i)\rangle\nonumber\\{}\nonumber\\
&=& {1\over\pi} {\pa\over\pa z}{\pa\over \pa z^{*}}~\langle {1\over N} ~\log~{\rm det}\left[(z-H)(z^*-H^\dgg)\right]\rangle
\eeqra
where we defined $z=x+iy$ (a careful derivation of (\ref{master}) is given for example in Section 2 of \cite{feinzee}.) It is at this point that the simplification associated with taking the large non-hermiticity limit may be appreciated: the determinant of $z-H$ for $H$ in (\ref{maxnh}) is simply
\beq\label{det}
{\rm det}~(z-H) = \left(\prod_{k=1}^N (z-w_i)\right) -1\,
\eeq
(For arbitrary $t$ and $h$ the corresponding formula is considerably more complicated.) Note that (\ref{det}) is completely symmetric in the $\{w_i\}$, and thus, for a given set of site energies it is independent of the way the impurities are 
arranged along the ring. Averaging over the impurities we obtain  \beq\label{avelog}
\langle {\rm log~det}~(z-H)\rangle = 2^{-N}\sum_{n=0}^{N} \left(\begin{array}{c}N\\n\end{array}\right)~{\rm  log}~\left[(z-r)^n (z+r)^{N-n} -1\right]\,.
\eeq

In the limit $N\rightarrow\infty$, the binomial coefficient appearing in (\ref{avelog}) is sharply peaked as a function of $n$ around $n\sim N/2$, and thus to first approximation,\footnote{We also get the correct normalization since in this approximation: $2^{-N} N!/((N/2)!)^2\sim 1$.}
\beq\label{thermo}
\langle {\rm log~det}~(z-H)\rangle \sim {\rm  log}~\left[(z-r)^{N\over 2} (z+r)^{N\over 2} -1\right]\,.
\eeq
In order to calculate the density of eigenvalues we may now insert
(\ref{thermo}) (and its counterpart for $H^{\dagger}$) into (\ref{master}), but due to the simplicity of (\ref{thermo}) we can avoid doing so and simply identify the branch singularities of the right hand side of (\ref{thermo}). 
These singularities clearly occur at 
\beq\label{singular}
z_n=\pm\sqrt{r^2 + e^{{4\pi i n\over N}}}\,,\quad n=0,1,\cdots,N/2\,,
\eeq 
which define the support of the density of eigenvalues.

To summarize, we see that the original unit circle spectrum of the 
deterministic part of (\ref{maxnh}) is distorted by the randomness (\ref{pw}) into the curve 
\beq\label{curve}
z^2=r^2 + e^{i\theta}\,,\quad 0\leq\theta<2\pi
\eeq
in the complex $z$ plane. Clearly, $r_c=1$ is 
a critical value of $r$. For $r<1$, the curve (\ref{curve}) is connected, whereas for $r>1$ it breaks into two disjoint symmetric lobes that are located to the right and to the left of the imaginary axis. The lobe on the right intersects the real axis at $E_{\rm min} = \sqrt{r^2-1}$ and at $E_{max} = \sqrt{r^2+1}$. As $r$ is decreased to $r_c=1$ the two lobes touch at the origin and merge into a single curve when $r$ becomes smaller than $r_c=1$.

In the three figures below we have plotted the curve (\ref{curve}) for three values of $r$ in the different regimes $r<1, r=r_c=1$ and $r>1$, on top
of the corresponding results of numerical simulations. 
\begin{figure}
\epsfysize=4 truein
\epsfxsize=4 truein
\centerline{\epsffile{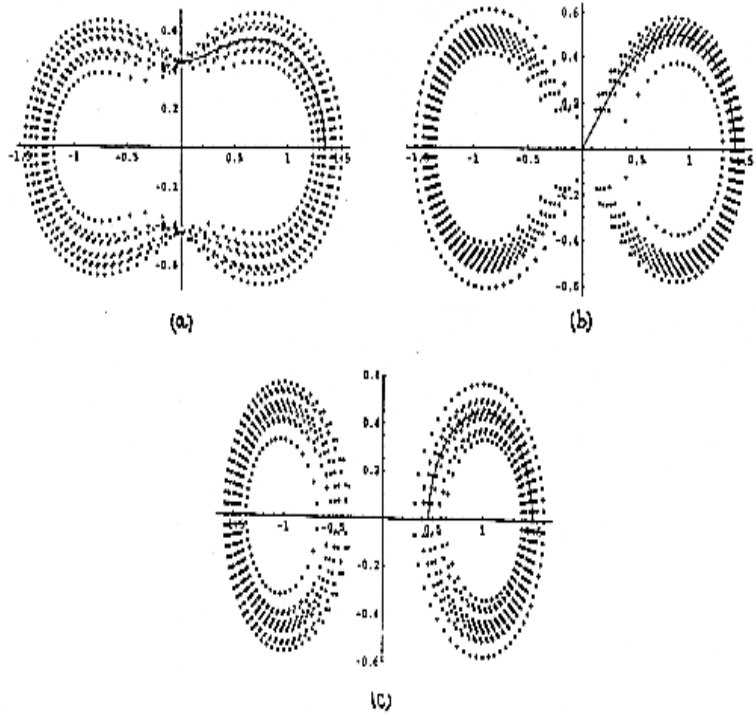}}
\vspace{-0.1 truein}
\caption[]{The spectrum of the Hamiltonian in Eq. (\ref{maxnh}). The site
energies are taken from (\ref{pw}) for various values of $r$. Figures (a), (b) and (c) correspond to $r=0.9, 1$ and $1.1$, respectively. The solid curve represents Eq. (\ref{curve}).}
\end{figure}

The width of the scatter of the numerical results around 
the analytical curve can be roughly estimated for any given $z_0$ on (\ref{curve}) by keepping (in addition to the leading term) all terms in the 
sum (\ref{avelog}) with $|n-(N/2)|\leq N\epsilon (z_0)$, where $\epsilon(z_0)$ is determined by steepest descent. Then the branch singularities 
on the right hand side of (\ref{avelog}) would have been bounded between the curves 
\beq\label{corrections}
(z^2-r^2)\left({z+r\over z-r}\right)^{\pm\epsilon} = e^{i\theta}\,.
\eeq

(Note added: In a forthcoming publication \cite{brz}, it is shown that the 
width of the scatter vanishes as $N\rightarrow\infty$ and that the spectrum 
of the Hamiltonian in (\ref{maxnh}) is actually one dimensional.)

\pagebreak

\section{Many impurities and critical transitions}
In this section we push the analysis of the previous section further and 
replace (\ref{pw}) by a generic probability distribution $P(w_i)$.

Using the determinant (\ref{det})
$\Delta(z)\equiv \left(\prod_{k=1}^N (z-w_i)\right) -1$ we write the Green's function for the ``one way" Hamiltonian (\ref{maxnh}) as
\beqast
G(z,z^*) = \langle {1\over N}~ {\pa_z\Delta(z)\over \Delta(z)}\rangle = 
\langle {1\over N}~ \sum_{k=0}^\infty\sum_{i=1}^N {(\Delta (z))^{-k}
\over z-w_i}\rangle\,.
\eeqast
Since each term in the previous equation factorizes, we finally arrive at the general formula
\beq\label{greens11}
G(z,z^*) = \langle {1\over z-w}\rangle - {1\over N}~ {\pa\over \pa z}~\sum_{k=1}^\infty ~{1\over k} \left[\langle \left({1\over z-w}\right)^k\rangle\right]^N\,,
\eeq
where in each term we average over $w$ against the arbitrary distribution $P(w)$. A simple check reveals that (\ref{greens11}) is consistent with (\ref{pw}) and (\ref{thermo}). The averages 
$\langle\left(z-w\right)^{-k}\rangle $ may be obtained of course by 
deriving the generating function 
\beq\label{gz}
g(z) =  \langle {1\over z-w}\rangle\,.
\eeq

It is worthwhile to mention here that a derivation of (\ref{greens11}) by expanding in powers of the ``one-way" hopping piece $H_0=\sum_{i}|i\rangle\langle i+1|$ in (\ref{maxnh}) has the nice feature that 
starting from any site $i$ on the chain, the terms that contribute to the trace (even without taking the average) are precisely those in which the particle hopped an integer number of complete revolutions around the chain (which is why 
the averages in (\ref{greens11}) are all raised to the power $N$.) This observation may provide a physical explanation of why the determinant (\ref{det}) is completely symmetric in the site energies $w_i$: the particle visits all sites equally as it hops, wherever the $w_i$'s are.

As a concrete application of (\ref{greens11}) we now concentrate on the Lorentzian  distribution (\ref{lorentz}) $P(w) = (\gam/\pi)~ (w^2+\gam^2)^{-1}$.
We have $g(z,z^*) = \left(z+i\gam~{\rm sgn~(Im}~z)\right)^{-1}$ and obviously
\beq\label{generating}
\langle \left({1\over z-w}\right)^k\rangle = \left({1\over z+i\gam~{\rm sgn~(Im}~z)}\right)^k = \left(g(z,z^*)\right)^k\,.
\eeq
Thus, using (\ref{greens11}) we obtain the exact result
\beq\label{glorentz1}
G(z,z^*) = {g(z,z^*)\over 1- \left(g(z,z^*)\right)^N} = {\left(z+i\gam~{\rm sgn~(Im}~z)\right)^{N-1}\over \left(z+i\gam~{\rm sgn~(Im}~z)\right)^N-1}\,.
\eeq
We would like now to calculate the density of eigenvalues. We first investigate $\rho(x,y)$ off the real axis. Using (\ref{generalrelation}) ($\rho(x,y) = (1/\pi)~(\pa/\pa z^*)~ G(z,z^*)$) we find 
\beqra\label{arcs}
\rho(x,y) &=& {1\over \pi}~(z\pm i\gam)^{N-1}~{\pa\over\pa z^*}~{1\over (z\pm i\gam)^N-1}\nonumber\\{}\nonumber\\ 
&=& {1\over N} \sum_{k=0}^{N-1} \delta^{(2)} (z +i\gam ~{\rm sgn~(Im}~z)-\om_k)\,,
\eeqra
where $\om_k=e^{2\pi ik/N}$, and $z\pm i\gam$ in the first line of (\ref{arcs})
correspond to $z$ being in the upper or lower half plane, respectively. The complex eigenvalues are thus equally spaced along the union of two arcs of a circle of radius one. The upper arc is that part of a semicircular arc of
a unit circle (centered at the origin) that remains in the upper half plane
after being pushed a distance $\gam$ downward along the imaginary axis. (The lower
arc is of course the mirror image of the upper arc.) Each of these arcs is thus of length ${\rm arc~cos}~(2\gam^2-1)<\pi$ and carries $n_{{\rm arc}} = (N/2\pi)~{\rm arc~cos}~(2\gam^2-1)~<~(N/2)$ eigenvalues. This means that the arcs exist only as 
long as $\gam < \gam_c =1$, which should be contrasted with the situation in (\ref{rholorentz}). There, the single impurity does not perturb the unit circle, which persists for all values of $\gam$. 

The rest of the eigenvalues, which do not have space to live on the arcs, must have snapped onto the real axis and formed ``wings". The eigenvalue density along the real axis is generated when $\pa/\pa z^*$ hits the step-functions $\theta (\pm{\rm Im}~z)$ in (\ref{glorentz1}) and we find that it is given by 
\beq\label{rhoreal}
\rho_{{\rm wing}}(x,y) = {1\over \pi}~\delta(y)~{\rm Im} {(x-i\gam)^{N-1}\over (x-i\gam)^N -1} = -{\delta(y)\over \pi}~\mu^{N-1}~ {\mu^N~{\rm sin}~\phi + {\rm sin}~(N-1)\phi\over \mu^{2N} - 2\mu^N~{\rm cos}~N\phi + 1}\,,
\eeq
where we have defined $x-i\gam \equiv \mu~e^{i\phi}$. 

In the large $N$ limit (\ref{rhoreal}) tends to a particularly simple form.
It is clear that this form depends on whether $\mu~<~\mu_c=1$ or $\mu~>~\mu_c$. For $\mu~>~1$
\beq\label{rhorealN}
\rho_{{\rm wing}}(x,y) = -{1\over \pi}~{{\rm sin}~\phi\over \mu}~\theta (\mu^2-1)~\delta (y) =  {\gam\over \pi}~{1\over x^2+\gam^2}~\theta (x^2+\gam^2-1)~\delta (y)
\eeq
(which again should be contrasted with the one impurity case
(\ref{rholorentz}),) while for $\mu~<~1$ $\rho_{\rm wing}=0$.

We thus see that for $\gam~<~\gam_c = 1$, there are two wings that bifurcate from the arcs at $x=\pm x_c = \pm \sqrt{1-\gam^2}$. Integrating over (\ref{rhorealN}) we find that the fraction of eigenvalues that reside in 
the wings is $(n_{\rm wings}/N) = 1-(2/\pi)~{\rm arc~cos}~\gam = 1-(1/\pi)~{\rm arc~cos}~(2\gam^2-1)$, which together with the fraction of eigenvalues $2(n_{{\rm arc}}/N) = (1/\pi)~{\rm arc~cos}~(2\gam^2-1)$ that reside
in the arcs, sum up exactly to $1$. As $\gam$ tends to $\gam_c$,
$x_c$ becomes smaller, and vanishes at $\gam=\gam_c$. At this point the two wings touch at the origin and the two arcs disappear completely.

As already mentioned, there is no critical $\gam_c$ in the single impurity 
case, as perhaps might be expected.  The many impurities case is dramatically
different: the long tail of the Lorentz distribution can overwhelm the
non-hermiticity for $\gam ~>~\gam_c$, and the spectrum collapses to the real axis.  We have carried out some numerical studies.  For $\gam$ away from $\gam_c=1$, our analytic results fit the numerical data closely, but as $\gam$ approaches $\gam_c$, the
statistical fluctuation between different realizations of $\{w_i\}$   becomes larger and larger (at $\gam =0.9$  for example, for $N$ as large as 400.)  It would be
interesting to study the character of the transition in more detail.

It is perhaps worthwhile to remark that although in many discussions of
disordered physics the gaussian distribution is the simplest to deal with,
here it leads to a $g(z)$ not given by an elementary function.

Having derived a closed formula (Eq. (\ref{greens11})) for the Green's function of the ``one-way" model (\ref{maxnh}) (with pure clockwise hopping $H_0=\sum_{i}|i\rangle\langle i+1|$), we may now perturb it by adding a term $\Delta H = \tau \sum_{i}|i+1\rangle\langle i|$ (with $\tau$ small) which allows the particle to hop counter-clockwise. To first order in perturbation theory, the correction to the Green's function (\ref{greens11}) 
is now
\beqast
\rmtr~\left[{1\over z-w}~\left(H_0~{1\over z-w}\right)^{N+1}~\left(\Delta H~{1\over z-w}\right)\right]
\eeqast
which reflects the fact that the particle has completed one revolution around the chain by performing
$N+1$ steps counter-clockwise and one step clockwise (which can occur
anywhere along the chain.) It is thus possible to study the non-hermitean Hamiltonian (\ref{h}) by perturbing both from the hermitean limit and the maximally non-hermitean limit.

In \cite{feinzee} we showed that by a hermitization method we can associate 
with every non-hermitean Hamiltonian $H$ a hermitean Hamiltonian ${\cal H}$. In particular, for the hopping Hamiltonians $H$ discussed in this paper, the Hamiltonian ${\cal H}$ involved the hopping of a particle with a binary internal state (call it an "up" or a "down" particle) such that as it hops it 
flips its internal state.  Solving the Schr\"odinger equation associated with
${\cal H}$ amounts to a simultaneous solution of the Schr\"odinger equations associated with the two hermitean positive Hamiltonians $H_1 = (z-H)(z-H)^\dgg
$ and $H_2 = (z-H)^\dgg (z-H)$.  We see that for our ``one-way" models 
$H_1$ and $H_2$ involve only nearest-neighbor hopping, while for the non-hermitean hopping models such as (\ref{h}) or its generalizations 
described in Section (5), $H_1$ and $H_2$ involve next-nearest neighbor 
hopping\footnote{This observation persists in the continuum of course.
In the continuum the ``one-way" Hamiltonian becomes a first order differential operator and thus $H_1$, $H_2$ are second order differential operators. The generic hopping Hamiltonian becomes a second order differential operator and thus $H_1$ and $H_2$ are fourth order differential operators.}

\pagebreak

\section{Continuum ``One Way" models}
The discrete ``One Way" models of the two previous sections were solved exactly in
the lagre $N$ limit. We now show that their continuum counterparts 
are also exactly solvable.

Starting with the continuum non-hermitean Schr\"odinger equation
\beq\label{hm}
H = - (1/m)[\pa_x + h]^2 + W(x)
\eeq
(with constant $h$), we reach the ``One Way" limit by letting both $h$ and $m$ tend to infinity 
with a finite ratio $h/m$ (which we set to $1/2$.) In this limit, $H$ in (\ref{hm})
turns into the first order non-hermitean operator
\beq\label{first}
H = -\pa_x + W(x)\,,
\eeq
which is the desired continuum ``One Way" model.
Clearly, the spectrum of (\ref{hm}) may be solved explicitly for any $W(x)$, once
the boundary conditions are specified. For example, if (\ref{first}) is defined
over $0\leq x\leq L$ with periodic boundary conditions, we have \beqra\label{firstpbc}
\psi_n(x)~ &=& {1\over\sqrt{L}}~{\rm exp}~(~\int\limits_0^x dy W(y) - E_n x)\nonumber\\
\phi_n^\dgg(x) &=& {1\over\sqrt{L}}~{\rm exp}~(-\int\limits_0^x dy W(y) + E_n x)
\eeqra
where $\psi_n$ and $\phi_n$ are the eigenvectors on the right and on the left, respectively. Imposing the boundary conditions we find
$E_n = {1\over L}~\int\limits_0^L dy W(y) + {2i\pi n\over L}$ ($n$ being 
an integer.)  Since we know the spectrum $\{E_n\}$ explicitly, we can now consider randomizing $W(x)$ in (\ref{first}) and calculate any
desired correlation function given any distribution of the random $W(x)$. For example, if $W(x)$ is drawn from the Gaussian distribution $P[W] = (1/Z)~{\rm exp}~[-(1/2g^2)\int_0^L dx W^2(x)]$ we readily find that the averaged density of eigenvalues is
\beqra\label{dos}
\rho({\rm Re}E, {\rm Im}E)&\equiv & \langle \sum_{n=-\infty}^{\infty} \delta\left({\rm Im}E-{2\pi n\over L}\right) \delta\left( {\rm Re}E - {1\over L}\int_0^L dx W(x)\right)\rangle\nonumber\\ &=& \sqrt{2\pi} 
(L/g) e^{-(L{\rm Re}E)^2/2g^2}\left[\sum_{n=-\infty}^{\infty} \delta\left({\rm Im}E-{2\pi n\over L}\right)\right]\,,
\eeqra
namely, each of the purely imaginary eigenvalues of $H_0=-\pa_x$, is smeared along the real axis by the fluctuations of the real potential $W(x)$. The spectrum
is thus one dimensional, as in the discrete case.

\pagebreak  

\section{Random hopping models and their mapping onto random walkers}
In this section we study a different class of models, in which there is no
site energy, but the hopping is random and non-hermitean. We consider the Schr\"odinger equation
\beq\label{steq}
E\psi_j=s_j^* \psi_{j+1}+t_{j-1} \psi_{j-1}
\eeq
with the hopping amplitudes $s_j$ and $t_j$ generated randomly according to
some prescription.

If the Hamiltonian is hermitean, then $s_j=t_j$. This hermitean problem was
studied some twenty years ago by Eggarter and Riedinger \cite{alsas}. They mapped the model onto a random walk problem and were able to show that all 
the states,
except for the one at $E=0$, are localized, and furthermore, that the
localization length diverges like $|{\rm log} E|$ as $E \rightarrow 0$. The
existence of localized states is in accordance with the arguments of
Anderson et al. \cite{anderson}. In contrast, the appearance of an extended state at
precisely $E=0$ is not generic and is due to the invariance of the spectrum under $E\rightarrow -E$ (as one can see by flipping the sign of $\psi_i$
for $i$ odd.) Thus, the extended state at $E=0$ is unstable under any perturbation that destroys this symmetry, such as adding random site energy.

Here we extend and generalize the analysis in \cite{alsas} to
the non-hermitean case. Our discussion below serves also as a review of \cite{alsas}, since obviously at any stage we can set $s_j$ equal to
$t_j$. Dividing (\ref{steq}) by $\psi_j$ and defining $\Delta_j \equiv
t_{j-1}\psi_{j-1}/\psi_j$ we obtain
\beq\label{delta}
\Delta_{j+1}={R_j \over E-\Delta_j }
\eeq
where $R_j \equiv s_j^* t_j$.
This equation is of course equivalent to Schr\"odinger's equation (\ref{steq}) and allows us to solve for $\Delta_j$ iteratively and hence for the wave function $\psi$. For a closed chain the obvious boundary condition is $\Delta_{N+1} = \Delta_1$. 

Performing a gauge transformation $\psi_j \rightarrow \lambda_j \psi_j$ we
find that we can effectively transform
$s_j^* \rightarrow (\lambda_{j+1}/\lambda_j) s_j^*$ and
$t_j \rightarrow (\lambda_j/\lambda_{j+1}) t_j$. As perhaps expected, $R_j$
is invariant under this transformation. We see by these considerations that the open chain and the
closed chain are quite different. For an open chain, we can always use this gauge freedom to set all the $t_j$ equal to $1$, say, with no loss of generality (which in the hermitean case would mean setting all the $s_j$ equal to $1$ as well, hence getting rid of randomness altogether.) On the other hand, for a closed chain with $N$ sites, we
have the boundary condition $\psi_{N+1}=\psi_1$ and hence the constraint $\lambda_{N+1}=\lambda_1$. The gauge transformation is not in general useful.

We can of course invent a variety of models. Hatano and Nelson \cite{hatano}
considered $s_j^*=\tau_j~e^\alpha$ and $t_j=\tau_j~e^{-\alpha}$ where $\tau_j$ is taken from the flat distribution  $P(\tau)= {1\over 2\Delta}~\theta(\Delta^2-\tau^2)$. We have studied one particularly simple 
model which we call the ``clock model", defined by setting $s_j$ and $t_j$ 
equal to random phases. In some sense, this model is particularly attractive, 
in that it includes no free parameters. Numerically, we found that the eigenvalues in the clock model are distributed (in what appears to be a uniform distribution) in a disk in the complex plane, centered at the origin and with radius approximately equal to $\pi/2$. The rotational invariance of 
the spectrum results from of our freedom to multiply the random 
Hamiltonian by an arbitrary overall phase. 

Recall that the spectrum of any non-hermitean (or hermitean) hopping problem as defined in (\ref{steq}) is invariant under $E\rightarrow -E$. Furthermore, in the case of the clock model, the existence of an extended state at $E=0$ (provided $E=0$ is in the spectrum, which is always the case when the number of sites $N$ is odd) survives the non-hermiticity. This is so because the Schr\"odinger equation (\ref{steq}) for $E=0$ implies that $|\psi_{j+1}|
=\Big | {t_{j-1}\over s_j^*}\psi_{j-1}\Big | = |\psi_{j-1}|$, and so this state is obviously extended. We expect the other states to be localized with an energy
dependent localization length that diverges as $E\rightarrow 0$. These expectations are supported by the numerical studies we have done. 
\begin{figure}
\epsfysize=4 truein
\epsfxsize=4 truein
\centerline{\epsffile{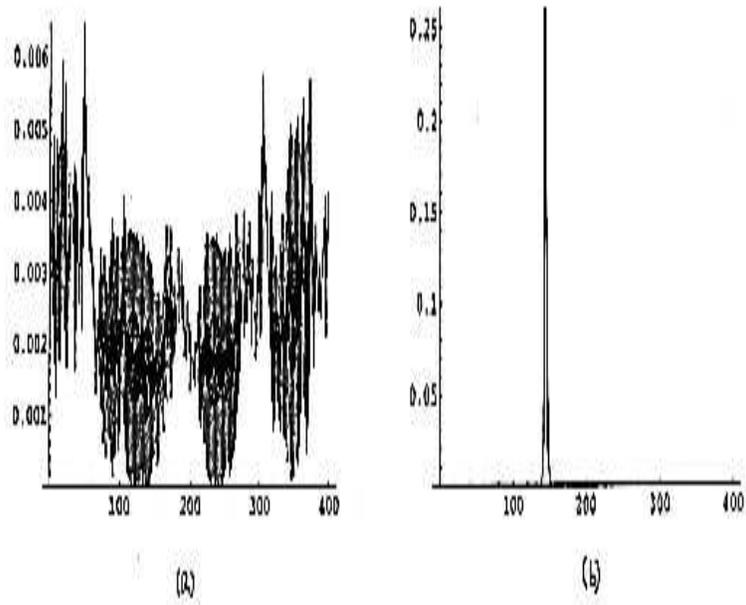}}
\vspace{-0.1 truein}
\caption[]{Two representative wave functions in one particular realization of the ``clock model". The wave function in Fig. (a) corresponds to the eigenvalue closest to the origin $E=0.04-0.14 i$. It is extended and its participation ratio is 0.75. The wave function in (b) corresponds to the eigenvalue farthest from origin
$E=-0.78+1.57 i$. It is localized and its participation ratio is 0.01.}
\end{figure}

We would like now to discuss the divergence of the localization length as $E\rightarrow 0$.  We thus focus on (\ref{steq}) for small $E$, that is $E$ small compared to a typical value of $\Delta_j$.  

In the hermitean case, we have the important observation that $R_j=|t_j|^2$
is real and positive. Also, $E$ is real, and thus by (\ref{delta}) we can take
$\Delta_j$ to be real (for an open chain, of course.) We see from (\ref{delta}) that for $E$ small compared to the typical scale of $\Delta$, the quantities $\Delta_j$ changes sign from site
to site, and hence, as pointed out by Eggarter and Riedinger, it is
convenient to iterate (\ref{delta}) twice and write
\beq\label{iterate}
\Delta_{j+2}= \left({R_{j+1}\over R_j}\right)~\left[{1-{E\over \Delta_j}\over 1+{E(\Delta_j-E)\over R_j}}\right]~\Delta_j\,.
\eeq
We can now define $z_j \equiv {\rm log} \Delta_j$ and interpret $z_j$ as the
position of a random walker in the complex plane and $j$ as time. Taking the logarithm of (\ref{iterate}) we obtain 
\beq\label{walker}
z_{j+2}=z_j + {\rm log}~{R_{j+1}\over R_j} + {\rm log}~\left[{1-{E\over \Delta_j}\over 1+{E(\Delta_j-E)\over R_j}}\right]
\eeq
defining the motion of the walker.

As noted above, for the hermitean case, $R_j$ is real and positive and thus the random walker stays on the real line. In particular, Eggarter and Riedlinger took $\langle {\rm log}~(R_{j+1}/R_j)\rangle=0$ and defined $\si^2\equiv 
\langle \left({\rm log}~(R_{j+1}/R_j)\right)^2\rangle$. They made the 
insightful observation that for $E\sim 0$ the last term in (\ref{walker}) (which would otherwise be too difficult to treat) could be taken into account effectively as boundary conditions set on the walker. Consider first the $-{\rm log}~\left[ 1+{E(\Delta_j-E)\over R_j}\right]$ part of that term. When $\Delta_j\sim (R_j/E)\sim$ a large positive number (we can always think of $\Delta_j$ as positive), the walker's position on the real line (namely, $x_j={\rm log}~\Delta_j$) decreases rapidly. We can thus effectively replace the term under consideration by a reflecting wall located at ${\rm log} (R^{{}^*
}/E)$ (which moves off to infinity as $E\rightarrow 0$), where $R^{{}^*
}$ is a typical value of $R_j$. Consider now the remaining ${\rm log}~\left[1-{E\over \Delta_j}\right]$ part of that term. This part becomes important when $\Delta_j={\cal O}(E)$, at which point the sequence of $\Delta_j$ switches sign.
We can thus think of a trap (or a ``walker-eating monster") located at $x\sim {\rm log} E$ (which wanders off to $-\infty$ as $E\rightarrow 0$.)
Eggarter and Reidliger showed that the localization length can be related to the lifetime of the walker. In a numerical simulation we start with a walker being born at the wall ($x\sim+\infty$); typically he drifts rapidly towards 
$x\sim 0$, where he executes a random walk in his middle age, and as soon as he drifts into a substantially negative $x$ region, he rapidly approaches his death at $x\sim -\infty$.

It is perhaps satisfying that in going from the hermitean problem to the non-hermitean problem the random walker has escaped from the one dimensional world and wandered off into the complex plane. (Strictly speaking, due to the properties of the logarithm, the walker now lives on a cylinder with circumference $2\pi$.) The equation (\ref{walker}) describing the two dimensional walker is considerably more difficult to treat.  One particularly simple case is given by the clock model in which the $\{R_j\}$ are pure random phases. Thus, separating $z_j=x_j+iy_j$, we see that as long as the term in square brackets in (\ref{walker}) is close to unity, the walker wanders in the $y$ direction, with its $x$ coordinate hardly changing.

We can estimate the localization length $L(E)$ rather crudely by noting that
in the (supposedly) exponential tail of the localized wave function $|\Delta_j|\sim {\rm exp}~(\pm 1/L(E))$ and so $x_j\sim \pm 1/L(E)$. Thus we estimate $1/L(E)$ to be given by some average $x$ coordinate
of the walker.  In numerical studies, we observe that indeed, with the
walker starting at $x=0$ he drifts into a random walk around some average
$<x>$, (but then eventually wanders off.)

One message of this section is that random (non-hermitean) hopping models
appear to be considerably more complicated than random (non-hermitean) site
energy models.  Indeed, let us mention that another interesting model we have studied, which we call a hopping ``sign model", is a restriction of the clock model, in which $s_j$ and $t_j$ are randomly (and independently) equal to $\pm 1$. 
This restriction destroys the rotational symmetry of the spectrum of the clock model, reducing it to a four fold symmetry, since eigenvalues must come in quadruplets $\pm E, \pm E^*$ (the Hamiltonian in this case is real and the symmetry of the spectrum 
under $E\rightarrow -E$ remains intact.) We obtained the density of eigenvalues numerically, and it exhibits a complicated interesting structure as shown in 
Fig.(4).
\begin{figure}
\epsfysize=4 truein
\epsfxsize=4 truein
\centerline{\epsffile{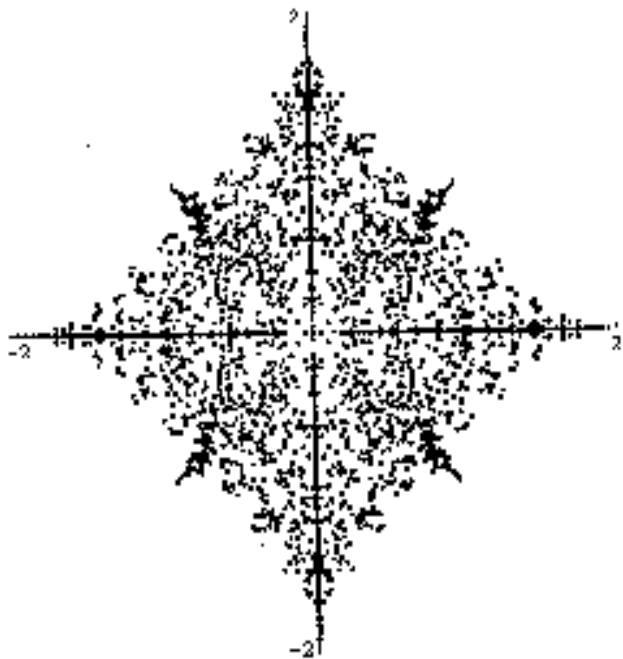}}
\vspace{-0.1 truein}
\caption[]{The spectrum of the hopping ``sign model" for a chain with 400 sites. The amplitudes $s$ and $t$ take on values $\pm1$ with equal probability.}
\end{figure}
It is an interesting challenge to calculate this structure analytically.

The hopping ``sign model" is the strictest restriction of the ``clock model". 
It is thus interesting to investigate how the eigenvalue distribution
of the ``sign model" changes when we allow the hopping amplitudes to take on more values. For example, we studied numerically a $Z_4$ model in which $s_j$ and $t_j$ took on (independently) values from $\{\pm 1, \pm i\}$ with equal probability. We found that the intricate structure in Fig. (4) was largely washed out towards the uniform disk distribution of the ``clock model", 
with some residual fine structure.

At this point we mention a class of hopping models in which the $s$'s and $t$'s in (\ref{steq}) are taken from the same probability 
distribution $P(x)$ such that $x$ is always real and positive.  Although the Hamiltonian $H$ is non-hermitean, the eigenvalues, that is, the solutions of det$(E-H) = 0$, are all real for $N$ large.  While surprising at first sight, this fact can be easily understood.  The crucial observation is that $R_j=s_j t_j$ are all real and positive.  For $N$ large, it does not make any difference whether the 
chain is closed or open as far as the eigenvalues are concerned, and so we can
apply the transformation $s_j\rightarrow s_j^{'} = (\lambda_{j+1}/\lambda_j) s_j$ and $t_j \rightarrow t_j^{'} = (\lambda_j/\lambda_{j+1}) t_j$  mentioned earlier 
to make
$s_j^{'}=t_j^{'}=\sqrt{R_j}$ (which is achieved by choosing $\lambda_{j+1} /\lambda_j =\sqrt{t_j / s_j}$.)  Thus the model is effectively hermitean, that is, the eigenvalues (which are gauge invariant, of course) are real. Note that the ``gauge" transformation just mentioned does not preserve the localization property of the wave function $\psi_j$. For instance, in numerical work, it is convenient to study the participation ratio, defined by\footnote{In general,
in addition to the eigenvector on the right corresponding to an eigenvalue $E$, $H\psi=E\psi$, there is also the eigenvector on the left, $H^\dgg \phi=E^*\phi$. With the normalization condition $\sum_j \phi_j^*\psi_j =1$, we can naturally consider the variant definition of the participation ratio given by 
$P(E)=1/(N\sum_j |\phi_j^*\psi_j|^2)$.} $P(E)=1/(N\sum_j |\psi_j|^4)$, such that $P\rightarrow 0$ for a localized state and $P\rightarrow 1$ for an extended
state. Clearly, $P$ is not preserved by the ``gauge" transformation in general,
and therefore, say, a localized state may be gauged into an extended state
(or vice-versa), unless the $\lambda_j$ are appropriately bounded as a
function of the site index $j$. We thus have to do a case by case study  
of how the $\lambda_j$'s behave as a function of $j$ to infer the localization properties of the original wave functions from the gauge transformed wave functions of the effectively hermitean problem, which are of course all localized, as predicted by Anderson and others.

Note that the ``sign model" mentioned in the previous paragraph
and the hopping Hamiltonian $H_0$ in (\ref{h0}) both evade the defining conditions of this class of crypto-real models.  An interesting question in mathematical physics is to calculate the fraction of eigenvalues escaping into the complex plane when the support of $P(x)$ includes negative values of $x$. 

Finally, in connection with the question raised in the last paragraph, 
we mention here a class of hopping models in which the $s$'s and $t$'s in (\ref{steq}) are of the form $s_j = t + T_j$ and $t_j = t - T_j$, where the independent random amplitudes $\{T_j\}$ take values in the range $-T\leq T_j \leq T$ according to some, say, even probability distribution (this model is a one dimensional discrete analog of the two dimensional ``model I" in \cite{miller}.) For $t^2>T^2$ all $R_j$ are clearly positive with probability one, and thus the model is ``crypto-real". In general, the  eigenvectors of the hermitean gauge transformed Hamiltonian would be all localized. Eigenvalues would start migrating into the compex plane only when $t^2<T^2$.

An amusing exception to the assertion that all states of the hermitean gauge transformed Hamiltonian are localized for $t^2>T^2$ is provided by the case in which the $\{T_j\}$ take on values $\pm T$ with equal probabilities. In this case $R_j\equiv R = t^2-T^2$ is actually deterministic and site independent and describes free hopping. Obviously, all states of the hermitean Hamiltonian are extended, regardless of the magnitude of disorder $T$. Clearly, $\sqrt{R}$, and thus all eigenvalues are real when $t^2>T^2$, and become pure imaginary for $t^2<T^2$. Thus, for this particular
sign distribution, the answer to the question raised above is that for $t^2<T^2$ all eigenvalues escape into the complex plane (and onto the imaginary axis.)

\pagebreak

\section{Conclusion}

We have studied localization and delocalization in a wide class of
non-hermitean Hamiltonians.  In various simplifying limits we were able to
obtain analytic expressions.  Our first model (\ref{gz1}), while it involved only a
single impurity, was able to capture the non-perturbative essence of the
localization-delocalization transition.  It was widely applicable to a
generic $H_0$, in any spatial dimensions.  For more explicit expressions, we
had to invoke the maximally non-hermitean, or ``one-way" limit, in which
the spectrum of $H_0$ was a circle.  (It is difficult to imagine that
comparably simple expressions can be obtained when the spectrum of $H_0$ is an
ellipse.) We were then able to analyze the problem of infinitely many
impurities, one at each site.  Results were given for several representative
probability distributions $P(w)$ for the impurity potential energy $w$.  We
analyzed the phase transitions as parameters in $P(w)$ were varied.

We studied non-hermitean random hopping Hamiltonians.  It has been known
that the hermitean one dimensional random hopping problem can be mapped 
into a random walk problem on the real line.  We found that in going from the hermitean to the non-hermitean problem, the random walker steps off the real axis and goes wandering off into the complex plane.  We showed by numerical 
work that for the random phase or ``clock model" the density of states 
appeared to be uniformly distributed over a disk, while in contrast for the ``random sign" model the density of states showed a fascinatingly intricate structure.

The study of non-hermitean Hamiltonians opens up a rich area for
exploration.  We can immediately think of many questions to be answered.
For instance, consider the many body problem (see also \cite{hatano}.) In the ``one-way" model of (\ref{maxnh}) the many (non-interacting)
fermion ground state will be described by a Slater determinant $\Psi \sim~{\rm det}_{i,j} [\psi_i(x_j)]$ which in the large $N$ limit assumes the form  $\prod_{i>j}~[e^{i\theta_i}-e^{i\theta_j}]$.  The question
arises, even before we consider impurities, on what is meant by the lowest
energy state of $H_0$. In other words, as we fill the system with non-interacting fermions,
how do we order the energy levels if they are arranged in a circle?  These
questions may perhaps be answered by coupling the system to some agent (such as
a radiation field or a heat bath) which can exchange energy with the many
body system.

As another set of questions, we can ask how a non-hermitean many body
system can be second quantized?  What are some of the properties of a
non-hermitean quantum field theory, such as a gauge theory in which the gauge
potential $A_\mu$ is non-hermitean? In a relativistic theory, if the 
Hamiltonian becomes non-hermitean,
does Poincar\'e invariance imply that its other nine generators should become
non-hermitean as well?

{\bf Acknowledgements}~~~
One of us (A. Zee) thanks D. Nelson for interesting him in this problem and
for continuing discussions, P. W. Anderson and M. P. A. Fisher for discussions
about localization, and J. Chalker for suggesting that we examine the
Lorentz distribution.  We also thank J. Miller for calling our attention to
\cite{miller}. Part of this work was done at the Institute for
Advanced Study where he was the Dyson Distinguished Visiting Professor and
he thanks the Dyson Fund for its support. This work was partly supported by the 
National Science Foundation under Grant No. PHY89-04035.


\newpage
\setcounter{equation}{0}
\renewcommand{\theequation}{A.\arabic{equation}}
{\bf Appendix : {Non-hermitean perturbation theory}}
\vskip 5mm

It is of course a trivial exercise to treat the model in (\ref{h})
perturbatively in the strength of the random impurity.  Denote the
eigenvalues of $H_0$ and $H$ by $\{E_\mu\}$ and $\{E_\mu(w)\}$          respectively.  Write $H_0 = S^{-1} E S$ with $E$ the diagonal matrix with elements $\{E_\mu\}$ (and thus the columns of $S^{-1}$
and the rows of $S$ are the right and left eigenvectors of $H_0$, respectively.)  Define
\beq\label{1}
E_\mu(w) = E_\mu + \sum_i w_i\,E_{\mu i} + \sum_{i,j} w_i w_j\,E_{\mu i j} + \cdots
\eeq
Then by simple arithmetic we obtain 
\beqra\label{2}
E_{\mu i} &=& S^{-1}_{i\mu}S_{\mu i}\nonumber\\
E_{\mu i j} &=& \sum_{\nu\neq\mu} { S^{-1}_{i\mu}S_{\mu j} S^{-1}_{j\nu}S_{\nu i}\over E_\mu-E_\nu}
\eeqra
and so on.
It follows that the averaged density of eigenvalues of $H$ is given to ${\cal O}(w^2)$ by
\beqra\label{3}
\rho(x,y) &=& \rho_0(x,y) + {w^2\over N} \sum_\mu\left\{ \left[ \delta^\prime (x-{\rm Re}~E_\mu)~\delta^\prime (y-{\rm Im}~E_\mu)~\sum_i {\rm Re}~E_{\mu i}~
{\rm Im}~E_{\mu i}\right]\right.\nonumber\\
&-&\left.{1\over 2}\left\{2\delta^\prime (x-{\rm Re}~E_\mu)~\delta (y-{\rm Im}~E_\mu)~\sum_i {\rm Re}~E_{\mu i i}\right.\right.\\ &-& \left.\left.\delta^{\prime\prime} (x-{\rm Re}~E_\mu)~\delta (y-{\rm Im}~E_\mu)~\sum_i ({\rm Re}~E_{\mu i})^2 + [x\leftrightarrow y, \, {\rm Re}\leftrightarrow {\rm Im}]\right\}\right\}\nonumber
\eeqra
where we have used only  $\langle w_i w_j\rangle = w^2\delta_{ij}$.             The result (\ref{3}) holds for all $H_0$ (and for finite $N$.)  For the 
specific $H_0$  in (\ref{h0}) and in the large $N$ limit there are drastic simplifications.  We have $E_{\mu i} = 1/N$ and so ${\rm Im~}E_{\mu_i} =0$            and $\sum_i \left( {\rm Re~} E_{\mu i}\right)^2 = 1/N$. Thus the only non-trivial quantity that comes in is 
\beq\label{4}
\sum_i E_{\mu i i} = {1\over N t} \sum_{\nu\neq\mu} {1\over {\rm cosh}~(h+ ik(\mu)) - {\rm cosh~} (h + ik(\nu))}
\eeq
where $k(\mu) = 2\pi\mu/N, \quad \mu = 0, 1, \cdots N-1$. The expression 
(\ref{3}) simplifies to 
\beq\label{5}
\rho(x,y) = {1\over N} \sum_\mu \delta(x- {\rm Re~}E_\mu - w^2 \sum_i {\rm Re~}E_{\mu i i}) \delta(y- {\rm Im~}E_\mu - w^2 \sum_i {\rm Im~}E_{\mu i i})\,.                \eeq
The sum in (\ref{5}) can be converted to an
integral and evaluated in principle.  For the ``one way" model described in
the text the integral is particularly simple and we obtain 
\beq\label{6}
\rho(x,y) = \int\limits_0^{2\pi} {d\theta\over 2\pi} ~\delta\left(x-(1+{w^2\over 2}) {\rm cos~}\theta\right)~\delta\left(y- (1-{w^2\over 2}) {\rm sin~}\theta\right)\,.
\eeq
To this order in $w$, the circle has been turned into an ellipse. Of course,
we do not see any trace of the wings in perturbation theory.

\end{document}